\documentstyle[psfig,prb,aps]{revtex}
\begin{document}
\draft

\twocolumn[\hsize\textwidth\columnwidth\hsize\csname
@twocolumnfalse\endcsname

\title{
Melting behavior of large disordered sodium clusters.
}
\author{Andr\'es Aguado}
\address{Departamento de F\'\i sica Te\'orica,
Universidad de Valladolid, Valladolid 47011, Spain}
%\date{submitted to Phys.~Rev.~B, July, 11, 1997}
\maketitle
\begin{abstract}
The melting-like transition in disordered sodium clusters
Na$_N$, with N=92 and 142 is studied by using a first-principles
constant-energy molecular dynamics simulation method.
Na$_{142}$, whose atoms are distributed in two (surface and inner) main shells
with different radial distances to the center of mass of the cluster,
melts in two steps: the first one,
at $\approx$ 130 K, is characterized by a high intrashell mobility of the atoms,
and the second, homogeneous melting, at
$\approx$ 270 K, involves diffusive motion of all the atoms across the whole
cluster volume (both intrashell and intershell displacements are allowed).
On the contrary, the melting of Na$_{92}$ proceeds gradually
over a very wide temperature interval, without any abrupt step visible in the
thermal or structural melting indicators. The occurrence of well defined steps
in the melting transition is then shown to be related to the existence of a 
distribution of the atoms in shells. Thereby we propose a necessary condition
for a cluster to be considered rigorously amorphouslike (totally disordered), 
namely that there are
no space regions of the cluster where the local value of the
atomic density is considerably reduced. Na$_{92}$ is the only cluster from the
two considered that verifies this condition, so its thermal behavior can be
considered as representative of that expected for amorphous clusters. 
Na$_{142}$, on the other hand, has a discernible atomic shell structure and
should be considered instead as just partially disordered. 
The thermal behavior of these two clusters is also compared to that of
icosahedral (totally ordered) sodium clusters of the same sizes.
\end{abstract}

\vskip2pc]

\section{Introduction}

Cluster melting is a topic of current theoretical
\cite{Bul92,Bla97,Ryt98,Cle98,Cal99,Agu99a,Agu99b} and experimental
\cite{Mar96,Sch97,Sch99,Hab99} interest,
motivated by the observation of several size-dependent properties which have
no analog in the bulk phase. Between those properties, we can mention the
following: (i) the melting like transition does not occur at a well defined
temperature as in the solid, but spreads over a finite temperature interval
that widens as the cluster size decreases. The lower end of that interval
defines the freezing temperature T$_f$ below which the cluster is completely
solidlike, their constituent atoms just vibrating about their equilibrium 
positions. The upper part of the interval defines the melting temperature
above which all the atoms can diffuse across the cluster and the ``liquid''
phase is completely established. Between those two temperatures the cluster
is not fully solid nor fully liquid.\cite{Jel86}
It is in that transition
region where the cluster-specific behavior emerges: (ia) Premelting effects
like partial melting of the cluster (the most usual case is surface melting)
\cite{Guv93}
or structural isomerizations upon heating, which lead to a melting in steps. 
\cite{Che96}
(ib) The dynamic coexistence regime, where the cluster can fluctuate in time
between being completely solid or liquid.\cite{Che91}
(ii) Strong nonmonotonic variations
of the melting temperature with size have been found in recent experiments
on sodium clusters. \cite{Sch97} The maxima in the melting temperature are not
in exact correspondence with 
either electronic or atomic shell closing numbers, but
bracketed by the two, suggesting that both effects are relevant to 
the melting process. It is important to note that the values of T$_f$ and T$_m$
as defined above are not yet amenable to the experiments, and that the
experimental melting temperature is somewhere between these two values. 

Previously\cite{Agu99a,Agu99b} we have reported density functional
orbital-free molecular dynamics (OFMD)
simulations of the melting process in sodium clusters
Na$_N$, with N=8,20,55,92, and 142. The OFMD technique\cite{Pea93}
is completely analogous to the method devised by Car and Parrinello (CP)
to perform dynamic simulations at an
{\em ab initio} level,\cite{Car85} but the electron density is taken as the
dynamic variable,\cite{Hoh64} as opposed to the Kohn-Sham (KS)
orbitals\cite{Koh65}
in the original CP method. This technique has been already used both in
solid state\cite{Sma94,Gov99} and cluster\cite{Bla97,Sha94,Gov95} physics.
In contrast to simulations which use empirical interatomic
potentials, the detailed electronic structure and the electronic contribution
to the energy and the forces on the ions are recalculated efficiently every
atomic time-step.
The main advantage over KS-based methods is that the computational effort
to update the electronic system
increases linearly with the cluster size N, in contrast to
the N$^3$ scaling of orbital-based methods. Indeed, these were the first
molecular dynamics simulations of large clusters as Na$_{92}$ and Na$_{142}$
that included an explicit treatment of the electronic degrees of freedom.

A very important issue in the simulations of cluster
melting is the election of the low-temperature isomer
to be heated. A good knowledge of the ground state structure (global minimum)
is required, as the details of the melting transition are known to be
isomer-dependent.\cite{Bon97} But the problem of performing a realistic
global optimization search is exponentially difficult as size increases, so
finding the global minima of Na$_{92}$ and Na$_{142}$ becomes impractical.
In our previous work\cite{Agu99b} we directly started from 
icosahedral isomers, as there is some experimental\cite{Mar96} and theoretical
\cite{Kum99} indications that suggest icosahedral packing in sodium clusters,
and found a good agreement with the experimental results of Haberland's
group.\cite{Sch97} However, we were unable to find those icosahedral structures
by an unconstrained search method as simulated annealing, which
always led to disordered isomers both for Na$_{92}$ and Na$_{142}$.
Although the icosahedral structures are more stable in all the cases, the energy
difference between both isomers is approximately 0.02 eV/atom, which is very
small. Amorphouslike structures have been found recently to be the ground state
isomers of gold clusters for a number of sizes,\cite{Gar98,Sol20} and pair 
potential calculations performed by Doye and Wales predict that the amorphous 
state is favored by long potential ranges.\cite{Doy96} The specific
features of those structures are little or no spatial symmetry and a pair
distribution function typical of glasses. Besides that, one usually finds a
large number of amorphouslike isomers nearly degenerate in energy, which 
suggests that they occupy a substantial fraction of the phase space available to
the cluster. Both the energy proximity to the more stable icosahedral isomers
and the large entropy associated with the amorphous part of the phase space
make plausible that amorphous isomers could be present in the experimental
cluster beams, so their thermal properties deserve specific investigation. 
Apart from this, the study of melting in amorphouslike
clusters is intrinsically interesting from a theoretical point of view.
Thus, the goals of this work are
to study the mechanisms by which the melting-like transition proceeds in
two disordered isomers of Na$_{92}$ and Na$_{142}$, to study the influence on
the melting behavior of the disorder degree, and to make a meaningful
comparison with the melting behavior of the corresponding icosahedral isomers.
In the next section we briefly present some technical details of the method.
The results are presented and discussed in section III and, finally,
section IV summarizes our main conclusions.

\section{Theory}

The orbital-free molecular dynamics method is a Car-Parrinello total
energy scheme\cite{Car85}
which uses an explicit kinetic-energy functional of the 
electron density, and has the electron
density as the dynamical variable, as opposed to the KS single particle
wavefunctions. 
The main features of the energy functional and
the calculation scheme have been described at length in previous work,
\cite{Pea93,Sha94,Bla97,Agu99a} and details of our method are as
described by Aguado et al.\cite{Agu99a} In brief, the electronic kinetic 
energy functional of the electron density, $n(\vec r)$, corresponds to 
the gradient expansion around the homogeneous limit through second order 
\cite{Hoh64,Mar83,Yan86,Per92}
\begin{equation}
T_s[n] =
T^{TF}[n] + {\frac{1}{9}} T^W[n],
\end{equation}
where the first term is the Thomas-Fermi functional (Hartree atomic units have
been used)
\begin{equation}
T^{TF}[n] = \frac{3}{10}(3\pi^2)^{2/3}\int n(\vec r)^{5/3}d\vec r,
\end{equation}
and the second is the lowest order gradient correction, where T$^W$,
the von Weizs\"acker term, is given by 
\begin{equation}
T^{W}[n] = 
\frac{1}{8}\int \frac{\mid \nabla n(\vec r) \mid^2}{n(\vec r)}d\vec r.
\end{equation}
The local density approximation is used for exchange and correlation.\cite
{Per81,Cep80} In the external field acting on the electrons,
$V_{ext}(\vec r) = \sum_n v(\vec r -\vec R_n)$, we take $v$ to be
the local pseudopotential of Fiolhais {\em et al.}, \cite{Fio95} which
reproduces well the properties of bulk sodium and has been shown to have good
transferability to sodium clusters.\cite{Nog96}
The cluster is placed in a unit cell of a cubic superlattice,
and the set of plane waves periodic in the superlattice
is used as a basis set to expand the valence density.
Following Car and Parrinello,\cite{Car85} the coefficients
of that expansion are regarded as generalized coordinates of a set of 
fictitious classical particles, and the corresponding Lagrange equations of 
motion for the ions and the electron density distribution are solved as 
described in Ref. \onlinecite{Agu99a}.

The calculations used a supercell of edge 
71 a.u. and the energy cut-off in the plane wave expansion of the 
density was 8 Ryd.
In all cases, a 64$\times$64$\times$64 grid was used. 
Previous tests \cite{Agu99a,Agu99b} indicate that the cut-offs used give good
convergence of bond lengths and binding energies. The fictitious
mass associated with the electron density coefficients
ranged between 1.0$\times$10$^8$ and 3.3$\times$10$^8$ a.u.,
and the equations of motion were integrated using the Verlet 
algorithm \cite{Ver65} for both electrons and ions with a time step 
ranging from $\Delta$t = 0.73 $\times$ 10$^{-15}$ sec. for the simulations 
performed at the lowest temperatures, to $\Delta$t = 0.34 $\times$ 
10$^{-15}$ sec. for those at the highest ones. These choices resulted in 
a conservation of the total energy better than 0.1 \%. 

Several molecular dynamics simulation runs at different constant energies 
were performed in order to obtain the caloric curve for each cluster.
The initial positions of the atoms for the first run were taken by slightly 
deforming the equilibrium low temperature geometry of the cluster.
The final configuration of each run served as the starting geometry for the
next run at a different energy. The initial velocities for every new run were
obtained by scaling the final velocities of the preceding run. The total 
simulation times varied between 8 and 18 ps for each run at constant energy.

A number of indicators to locate the melting-like transition were employed.
Namely, the specific heat defined by \cite{Sug91}
\begin{equation}
C_v = [N - N(1-\frac{2}{3N-6})<E_{kin}>_t ~ <E_{kin}^{-1}>_t]^{-1},
\end{equation} 
where N is the number of atoms and $<>_t$ indicates the average along a
trajectory;
the time evolution of the distance between each atom
and the center of mass of the cluster 
\begin{equation}
r_i(t) = \mid \vec R_i(t) - \vec R_{cm}(t)\mid;
\end{equation} 
the average over a whole dynamical trajectory of the 
radial atomic distribution function g(r), defined by
\begin{equation}
dN_{at} = g(r)dr
\end{equation} 
where $dN_{at}(r)$ is the number of
atoms at distances from the center of mass between r and r + dr;
and finally, the diffusion coefficient
\begin{equation}
D = \frac{1}{6}\frac{d}{dt}(<r^2(t)>),
\end{equation}
which is obtained from the long time behavior of the mean square
displacement $<r^2(t)>=\frac{1}{Nn_t}\sum_{j=1}^{n_t}\sum_{i=1}^N
[\vec R_i(t_{0j}+t)-\vec R_i(t_{0j})]^2$, where $n_t$ is the number of time
origins, $t_{0j}$, considered along a trajectory.

\section{Results}

The most stable disordered structures of Na$_{92}$ and Na$_{142}$ that we have
found with the simulated annealing technique are shown in Fig. 1, together with
the relaxed icosahedral isomers obtained as explained in our previous work.
\cite{Agu99b} The dynamical annealing runs\cite{Car85}
were started from high-temperature liquid
isomers thermalized during 30 ps at 600 K. The cooling strategy was to reduce
the internal cluster temperature by a factor of 0.9999 
of its actual value each twelve atomic
time steps. With the chosen time step of 0.34 $\times$ 10$^{-15}$ sec. the
temperature reduction is applied each four femtoseconds.
A first important difference between the icosahedral
and disordered isomers is that the former ones have a smoother surface. Besides
that, no apparent spatial symmetry is observed in the disordered isomers. In
Fig. 2 we show
the short-time averages (sta) of the distances between each atom and the 
center of mass of the cluster for both isomers of Na$_{142}$, obtained from
an OFMD run at a low
temperature. The values of $<r_i(t)>_{sta}$ are almost independent of time.
The clusters are solid, the atoms 
merely vibrating around their equilibrium positions. Curve crossings are due to
oscillatory motion and slight structural relaxations rather than 
diffusive motion. The difference between
Figs. 2a and 2b is due to structure. For the icosahedral isomer in Fig. 2a one 
can distinguish quasidegenerate groups 
which are characteristic of the symmetry: one line near the center of mass 
of the cluster identifies the central atom, whose position does not exactly 
coincide with the center of mass because of the location of the five surface 
vacancies (147 atoms are needed to form a complete three-shell icosahedron); 
12 lines correspond to the first icosahedral shell; another 42
complete the second shell, within which we can distinguish the 12 vertex atoms 
from the rest because their distances to the center of mass are slightly 
larger; finally, 82 lines describe the external shell, where again we 
can distinguish the 7 vertex atoms from the rest. In contrast, the lines for
the disordered isomer in Fig. 2b are quite dispersed. Nevertheless, there is a
narrow interval where the ionic density is very low, that serves to
define separate surface and inner atomic shells. The case of Na$_{92}$ 
(not shown) is similar, but in that case the structure is more uniformously
amorphous, and there is no way to distinguish between surface and inner shells.
We will see below that this small difference is very important.
Soler {\em at al.}\cite{Gar98,Sol20} have compared the structures of 
icosahedral and amorphous gold
clusters, with similar results: while the atoms in the icosahedral isomers are
clearly distributed in atomic shells with different radial distances to the
center of mass of the cluster, in the case of amorphous clusters there are
``atomic transfers'' between shells that blur the atomic shell structure.
In the case of gold clusters, the amorphous isomers turn out to be the 
minimum energy structures for a number of sizes.\cite{Gar98}
In the case of Na$_{92}$ and Na$_{142}$,
the icosahedral isomers are more stable than the lowest energy disordered
isomers found
(0.017 eV/atom and 0.020 eV/atom for Na$_{92}$ and Na$_{142}$ respectively).
\cite{Agu99b}

For each cluster we have calculated the internal temperature, defined as the
average of the ionic kinetic energy,\cite{Sug91}
as a function of the total energy--
the so-called caloric curve. A thermal phase transition 
is indicated in the caloric curve by a change of slope, the slope being the 
specific heat; the height of the step gives an estimate of the latent heat 
of fusion.
However, melting processes are more easily recognized as peaks in the specific 
heat as a function of temperature, that has been 
calculated directly from eq. (4).
The fact that the specific heat peaks occur at the same
temperatures as the slope changes of the caloric curve (see curves below) gives
us confidence on the convergence of our results, as those two quantities have
been obtained in different ways.

The specific heat curves for Na$_{142}$ (fig. 3) display two main
peaks, suggestive of a two-step melting process.
For the disordered cluster the two peaks are well
separated in temperature, T$^{dis}_s \approx$ 130 K and T$^{dis}_m \approx$ 
270 K, whereas they are much closer together for the icosahedral cluster,
T$^{ico}_s \approx$ 240 K and T$^{ico}_m \approx$ 270 K, so close
that only one slope change in the caloric curve can be distinguished
in this case. The results suggest that the melting transition
starts at a temperature T$_s$ and finishes at T$_m$ with the difference
in the melting of the two isomers being the smaller T$_s$ value 
of the disordered isomer. In our previous work\cite{Agu99b} we showed that
for the icosahedral cluster
those two steps are due to surface melting and homogeneous melting,
respectively. Here we show that a similar explanation is valid for the 
disordered Na$_{142}$ isomer.
At T=160 K, a temperature between T$_s^{dis}$ and T$_m^{dis}$,
the structure of the disordered cluster
is more fluid than at low temperature (fig. 2). 
Fig. 4 indicates that the spherical atomic shells have separately melted,
atoms undergoing diffusive motions mainly across a given shell, as seen in the
bold lines, without an appreciable intershell diffusion (although some 
occasional intershell displacement has been observed). The larger spread of
the upper bold line indicates that diffusion is
appreciably faster in the surface shell. Thus the transition at 130 K
can be identified with intrashell melting. Why it
occurs at a rather low temperature is associated with the large spread in the
radial distances of the atoms in each shell. The atomic shells
are structurally disordered at low temperature, although
kinetically frozen, like in a typical glass, so inducing diffusion in the
shells of that cluster is rather easy and occurs at moderate temperatures.
The surface melting stage does not develop in the icosahedral isomer
until a temperature of T$_s^{ico}\approx$ 240 K is reached. 
At this temperature, the time evolution of the surface 
atomic distances to the cluster center
becomes very similar to those of the disordered Na$_{142}$ isomer
at 160 K.\cite{Agu99b} 
Inducing diffusion in the surface of the icosahedral isomer
requires a higher temperature because of the higher structural order of its 
surface. Once the surface of both isomers has melted, homogeneous melting
occur at the same temperature, $\approx$ 270 K, in very good agreement with
the experimental value of 280 K.\cite{Sch97} From that temperature on, the
liquid phase is completely established (all atoms can diffuse across the
whole cluster volume)\cite{Agu99b} 
and differences between both isomers have no sense anymore.

The thermal behavior of both isomers is not so different. 
The radial atomic density distribution of the disordered isomer at 160 K
(Fig. 5) shows a smoothed shape
with respect to that found at low T, but a distribution of the atoms in 
separate surface
and inner atomic shells can be clearly distinguished. In fact, the intermediate
temperature distribution is similar to that found for the icosahedral
isomer in the same temperature region,\cite{Agu99b} 
as the atomic shells are equally distinguished. The small
gap in the ionic density (Figs. 2 and 5) of the disordered isomer at low
temperature drives the cluster towards a well defined shell structure upon
heating. We conclude that, despite the high orientational disorder in both 
surface and
inner shells, the cluster can not be considered completely amorphous, as Fig. 5
at intermediate temperature shows some radial atomic ordering.
There are still more similarities. The 
solidlike phase of the icosahedral isomer disappears as soon as a temperature
of 130 K is reached, even though no peak in the specific heat is detected:
there are isomerization transitions between different 
permutational isomers which preserve the icosahedral structure.\cite{Agu99b}
These isomerizations involve the displacement of the five surface vacancies
across the outer shell.
Thus, both isomers leave the solidlike phase at $\approx$ 130 K,
the only difference being that one has direct access to the intrashell melting
stage while the other enters first an isomerization regime. Calvo and
Spiegelmann\cite{Cal99} have related the appareance of specific heat peaks to
sudden openings of the available phase space. In the isomerization regime the
icosahedral cluster has access just to a limited number of symmetry-equivalent
isomers, while the phase structure of the disordered cluster opens suddenly
to include a very large number of isomers, as all the possible position
interchanges between two atoms of a given shell are allowed. 
Thus, a specific heat peak appears at T$\approx$130 K for the disordered 
isomer, but not for the icosahedral isomer.
Any atomic shell distribution in the time average of g(r) 
disappears completely when homogeneous melting occur (Fig. 5).

The results for Na$_{92}$ are shown in fig. 6.
Two-step melting is again observed in the icosahedral isomer, with a small
prepeak in the specific heat at 
T$_s^{ico} \approx$ 130 K and a large peak, corresponding to
homogeneous melting, at T$_m^{ico} \approx$ 240 K. In this
case T$_s^{ico}$ and T$_m^{ico}$ are well separated. T$_s^{ico}$
is in the range where the isomerization processes in icosahedral Na$_{142}$
set in and the intrashell melting stage in disordered Na$_{142}$ develops. 
The larger number of vacancies in the surface shell of icosahedral Na$_{92}$,
as compared to icosahedral Na$_{142}$,
allows for true surface diffusion and these processes give rise to a distinct
peak in the specific heat.\cite{Agu99b} The disordered isomer melts gradually
over a wide temperature interval, and no prominent specific heat peaks nor
important slope changes in the caloric curve are detected. Ercolessi 
{\em et al.}\cite{Erc91} have also found a melting process without a latent heat
of fusion for amorphous gold clusters with less than $\sim$ 90 atoms. In Fig. 7
we show the radial ionic density distribution of disordered Na$_{92}$ at several
representative temperatures. At a temperature as low as 70 K there is no
discernible atomic shell structure. The g(r) function for a higher temperature
where the surface of the icosahedral isomer has already melted is not very
different. At a temperature where the icosahedral isomer is liquid the only
appreciable change in g(r) is due to the thermal expansion of the cluster. The
structure of cold disordered Na$_{92}$ is both radially and orientationally
disordered.
The structure is kinetically frozen, but there seems to be no barriers impeding
the exploration of the liquid part of the phase space. In fact, the cluster
is already in that region at low temperature, as Fig. 7 shows. 
This is seen most clearly in
the evolution of the diffusion behavior with temperature. In Fig. 8 we show
the temperature variation of the square root of the diffusion coefficient. While
the two steps in the melting of the icosahedral isomer are clearly detected
as slope changes at the corresponding temperatures, the value of $\sqrt{D}$
for the disordered isomer increases with temperature in a smooth way, without 
any abrupt change. Thus, the opening of the available phase space proceeds in a
gradual way and specific heat peaks are not detected.

We have found a very different thermal
behavior for two clusters that were classified in principle as disordered.
Although just two examples are not enough to draw general conclusions,
we believe that the thermal behavior typical of amorphouslike sodium
clusters is that found for Na$_{92}$, and that what is lacking is
a clear definition of what an amorphous cluster is. In line with the discussion
of ``atomic transfers'' between shells advanced by Soler {\em et al,}
\cite{Sol20} we propose that a large orientational disorder is not enough for
a cluster to be classified as amorphouslike. Only when
those atomic transfers are maximal, in such a way that no local regions with
low atomic density exist, the cluster can be considered completely amorphous. 
The existence of those regions, however small they
may be (as is the case of Na$_{142}$), promote the creation of appreciable
free energy barriers in the potential energy surface, so sudden access to a
substantial region of the available phase space is expected above a certain
temperature, and peaks will appear in the temperature evolution of the
specific heat. On the contrary, the absence of such low atomic density regions
facilitates the diffusion of the atoms across the whole cluster volume right
from the start of the heating process. As the liquidlike phase is established
precisely when all the atoms in the cluster can diffuse across the cluster
volume, we expect that no appreciable free energy barriers will be found in
these cases, and no specific heat peaks will be detected.

A few comments regarding the quality of the simulations and of the annealing
runs are perhaps in order here.
The orbital-free representation of the atomic interactions is much more
efficient than the more accurate KS treatments, but is still
substantially more expensive
computationally than a simulation using
phenomenological many-body potentials. Such potentials contain
a number of parameters that are usually chosen by fitting some bulk and/or
molecular properties. In contrast
our model is free of external parameters, although there are
approximations in the kinetic and
exchange-correlation functionals. 
The orbital-free scheme accounts, albeit approximately, for the effects of the
detailed electronic distribution on the total energy and the forces on the ions.
We feel that this is particularly important in metallic clusters for which a
large proportion of atoms are on the surface and experience a very different
electronic environment than an atom in the interior. Furthermore, the
adjustment of the electronic structure and consequently the energy and forces
to rearrangements of the ions is also taken into account.
But the price to be paid
for the more accurate description of the interactions is a less complete
statistical sampling of the phase space. The simulation times are substantially
shorter than those that can be achieved in phenomenological simulations.
The cooling rate employed in the annealing runs is also faster than those that
can be achieved by using empirical potentials. Nevertheless,
we expect that the
locations of the several transitions are reliable, because 
all the indicators we have
used, both thermal and structural ones, are in essential agreement on
the temperature at which the transitions start. On the other hand, 
the disordered isomers found in
different annealing runs did not show substantial structural or energetic
differences with respect to those studied here.

\section{Summary}

The melting-like transition in disordered Na$_{142}$ and Na$_{92}$ has been 
investigated by applying an orbital-free, density-functional molecular
dynamics method. The computational effort which is required is modest in 
comparison with the traditional Car-Parrinello Molecular Dynamics technique 
based on Kohn-Sham orbitals, that would be very costly for clusters of this
size. This saving allows the study of large clusters.

A disordered isomer of Na$_{142}$ melts in two steps as evidenced
by the thermal indicators. The transition
at T$_s^{dis} \approx$ 130 K
is best described as intrashell melting. This is followed at T$_m^{dis} 
\approx$ 270 K by homogeneous melting. Melting is found to depend on the 
starting low-temperature isomer. Specifically, 
for an icosahedral Na$_{142}$ isomer, the analysis of thermal, 
macroscopic properties places those two stages much closer in 
temperature, 240 K and 270 K respectively.\cite{Agu99b}
Nevertheless, isomerization transitions are observed in the icosahedral isomer
at a temperature as low as T$_s^{dis}\approx$130 K. These isomerizations involve
the motion of the five atomic vacancies in the cluster surface,
preserve the icosahedral structure and do not give rise to any 
pronounced feature in the caloric curve. In the disordered isomer
there is not a separate isomerization regime 
(something rather evident because there is not an underlying ordered structure
in each shell),
but there is a melting-in-steps process, due to the distribution of the atoms
in different shells.
Thus, the melting of both isomers is not as different as suggested by the
thermal indicators. An icosahedral isomer of Na$_{92}$ melts also in a similar 
way: there is a minor peak in C$_v$ at T$_s^{ico} \approx$130K that indicates 
surface melting, and a main, homogeneous melting peak at T$_m^{ico}\approx$240 
K. The lower value of T$_s^{ico}$, as compared to Na$_{142}$, is due to the 
larger number of surface vacancies, 
and the melting-in-step process is due to the atomic shell
structure. The melting of disordered Na$_{92}$ proceeds instead gradually, and
spreads over a very wide temperature interval. The thermal indicators as the
caloric curve or the specific heat do not show any indications of an abrupt
transition, which suggests that the phase space available to the cluster does
not increases suddenly at any given temperature. 
The square root of the diffusion coefficient increases with temperature in a
smooth way, in contrast to the step diffusive behavior of icosahedral 
Na$_{92}$. It has been suggested that the absence of any abrupt transition is
closely related to the absence of any shell structure in the radial atomic
density distribution, which should be considered a necessary condition for a
cluster to be considered completely amorphous. In this, sense, only the
disordered isomer of Na$_{92}$ can be considered rigorously amorphous, while
the disordered Na$_{142}$ isomer should be considered just partially amorphous.

In summary, we have found a number of structural properties that have an
important effect on the melting properties of sodium clusters. A melting in
steps process is to be expected in almost all clusters where 
a clear distribution of the atoms in radial shells exists, as is the case of
both isomers of Na$_{142}$ and of the icosahedral isomer of Na$_{92}$. In 
those cases, intrashell diffusive motion starts at a temperature 
T$_{intra}$, lower than the temperature T$_{inter}$ at which intershell 
diffusive motion begins to be important. The difference T$_{inter}$-T$_{intra}$
is small if the orientational order of the shells is large (for example,
icosahedral order) and the number of
vacancies in each shell is small: this occurs for icosahedral Na$_{142}$, with
just five surface vacancies; A limit case is icosahedral Na$_{55}$. With two
complete atomic shells, intrashell motions are as difficult as intershell
motions and the two transitions merge into one.\cite{Agu99b} When one shell
contains a large number of vacancies, the two temperatures are well separated
no matter how high the orientational order is: this is exemplified by the case
of icosahedral Na$_{92}$. Also, if the shells have a high orientational
disorder, the two transitions are well separated in temperature no matter how
close we are from an icosahedral shell closing: an example is the disordered
Na$_{142}$ isomer. Finally, a gradual melting process
without any abrupt transition is
to be expected for all those clusters which have both orientational and radial
disorder, that is for amorphouslike clusters: this is the case of amorphous
Na$_{92}$.

$\;$

$\;$

$\;$

{\bf ACKNOWLEDGMENTS:} 
This work has been supported by DGES 
(Grant PB98-0368)
and Junta de Castilla y Le\'on (VA70/99).
The author acknowledges useful discussions with J. M. L\'opez, J. A. Alonso,
and M. J. Stott.

%\newpage

{\bf Captions of Figures.}

{\bf Figure 1} Structures of the low temperature (a) amorphous Na$_{142}$,
(b) icosahedral Na$_{142}$, (c) amorphous Na$_{92}$ and (d) icosahedral 
Na$_{92}$ isomers.

{\bf Figure 2} Short-time averaged distances $<r_i(t)>_{sta}$ between each atom
and the center of mass in Na$_{142}$, as functions of time for 
(a) the icosahedral isomer at T= 30 K and (b) the amorphous isomer at T= 47 K.

{\bf Figure 3} Caloric and specific heat curves
of Na$_{142}$, taking the
internal cluster temperature as the independent variable. 
The deviation around the mean
temperature is smaller than the size of the circles.

{\bf Figure 4} Short-time averaged distances $<r_i(t)>_{sta}$ between each atom
and the center of mass in amorphous Na$_{142}$
as functions of time
at T= 160 K. The bold lines follow the 
evolution of a particular atom 
in the surface shell and another in the outermost core 
shell.

{\bf Figure 5} Time
averaged radial atomic density distribution of the
amorphous isomer of 
Na$_{142}$, at some representative
temperatures.

{\bf Figure 6} Caloric and specific heat curves
of Na$_{92}$, taking the
internal cluster temperature as the independent variable. 
The deviation around the mean
temperature is smaller than the size of the circles.

{\bf Figure 7} Time
averaged radial atomic density distribution of the
amorphous isomer of 
Na$_{92}$, at some representative
temperatures.

{\bf Figure 8} Square root of the diffusion coefficient as 
a function of temperature for the icosahedral and
amorphous isomers of Na$_{92}$.

\onecolumn[\hsize\textwidth\columnwidth\hsize\csname
@onecolumnfalse\endcsname

\begin{figure}
%\vspace{-10mm}
\psfig{figure=fig1a.epsi}
%\label{fig:ionization}
\end{figure}

\begin{figure}
%\vspace{-10mm}
\psfig{figure=fig1b.epsi}
%\label{fig:ionization}
\end{figure}

\begin{figure}
%\vspace{-10mm}
\psfig{figure=fig1c.epsi}
%\label{fig:ionization}
\end{figure}

\begin{figure}
%\vspace{-10mm}
\psfig{figure=fig1d.epsi}
%\label{fig:ionization}
\end{figure}

\begin{figure}
%\vspace{-10mm}
\psfig{figure=fig2a.epsi}
%\label{fig:ionization}
\end{figure}

\begin{figure}
%\vspace{-10mm}
\psfig{figure=fig2b.epsi}
%\label{fig:ionization}
\end{figure}

\begin{figure}
%\vspace{-10mm}
\psfig{figure=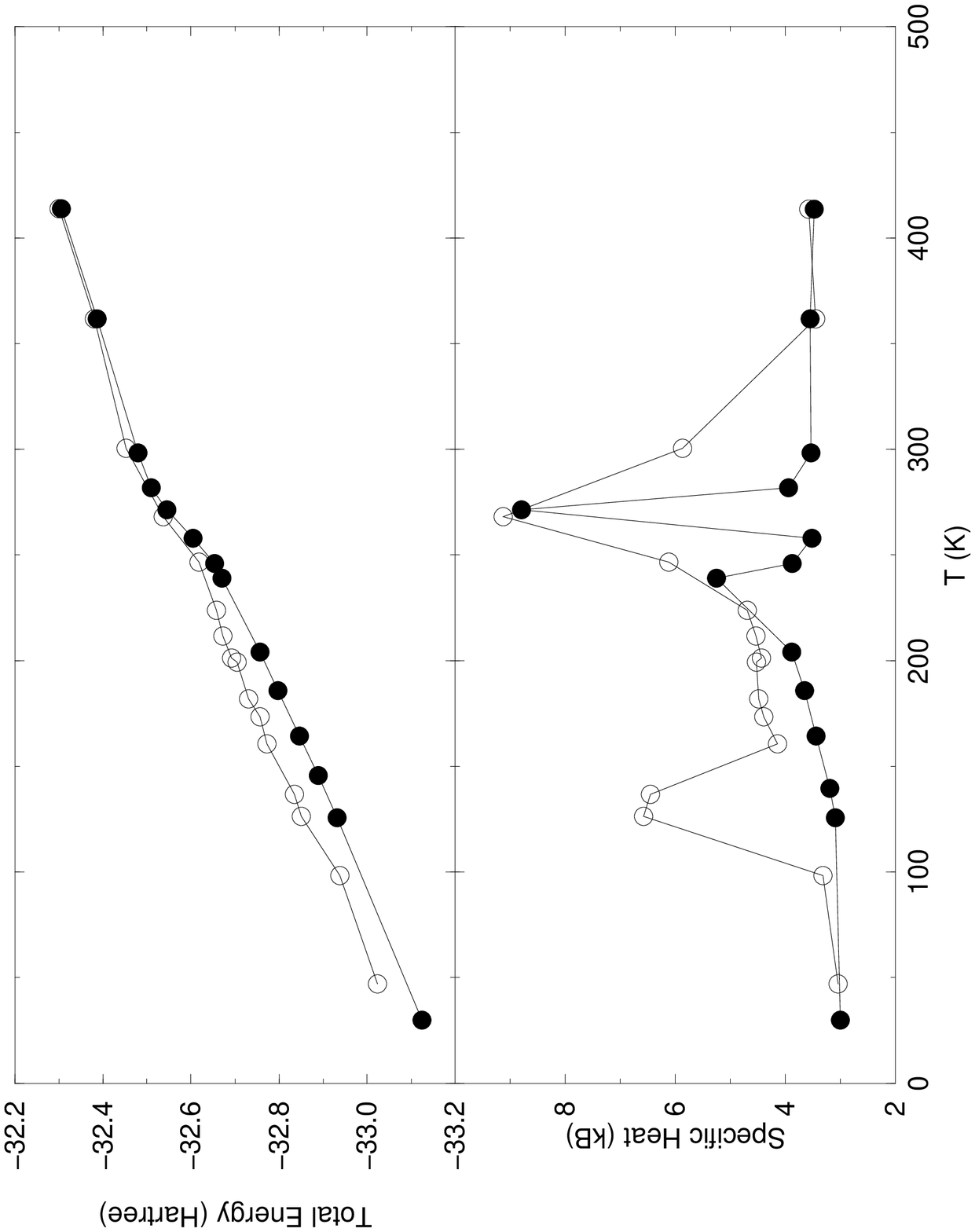}
%\label{fig:ionization}
\end{figure}

\begin{figure}
%\vspace{-10mm}
\psfig{figure=fig4.epsi}
%\label{fig:ionization}
\end{figure}

\begin{figure}
%\vspace{-10mm}
\psfig{figure=fig5.epsi}
%\label{fig:ionization}
\end{figure}

\begin{figure}
%\vspace{-10mm}
\psfig{figure=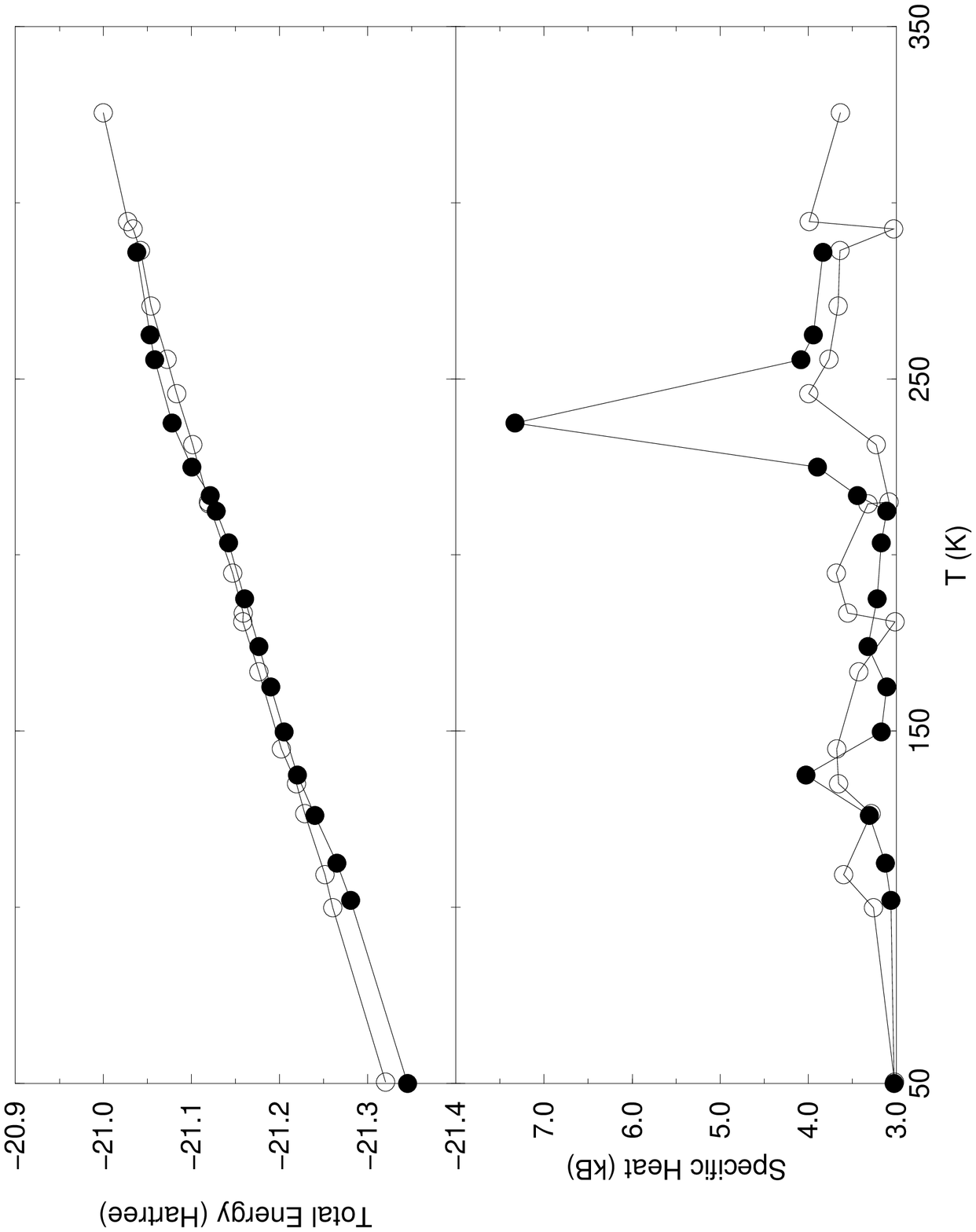}
%\label{fig:ionization}
\end{figure}

\begin{figure}
%\vspace{-10mm}
\psfig{figure=fig7.epsi}
%\label{fig:ionization}
\end{figure}

\begin{figure}
%\vspace{-10mm}
\psfig{figure=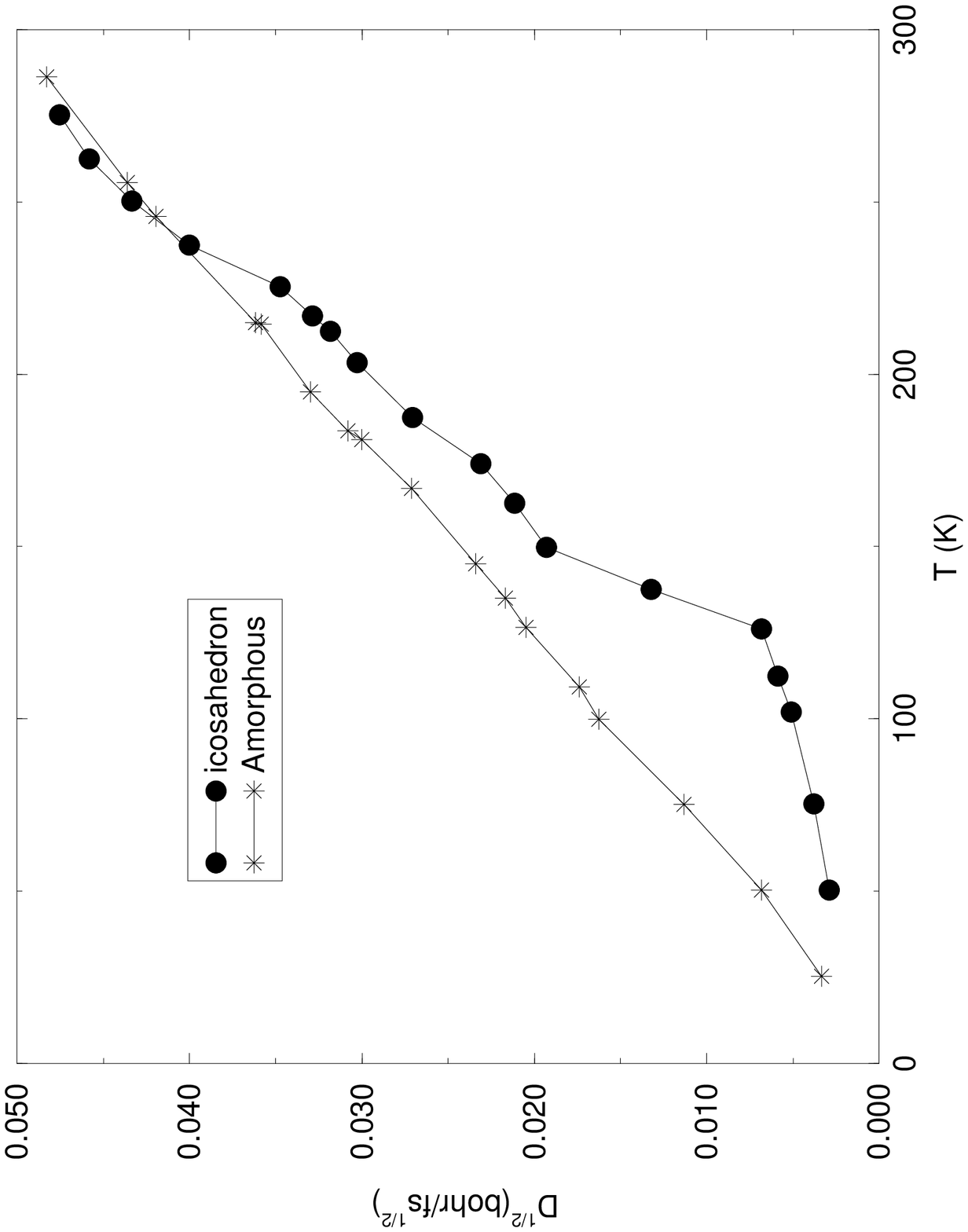}
%\label{fig:ionization}
\end{figure}


\begin{references}
%\bibitem[*]{bockstedte}{Present address:
%Lehrstuhl f\"ur Theoretische Fest\-k\"or\-per\-phy\-sik,
%Universit\"at Erlangen, Staudtstrasse 7/B2, 
%D-91058 Erlangen, Germany.
%}
%\bibitem[\dagger]{pehlke}{Present address:
%Physik Department T30, Universit\"at M\"unchen, James-Franck-Strasse, 
%D-85747 Garching, Germany.
%}
\bibitem{Bul92}{
A.~Bulgac and D.~Kusnezov, 
Phys.~Rev.~Lett. {\bf 68}, 1335 (1992);
Phys.~Rev.~B {\bf 45}, 1988 (1992);
N. Ju and A. Bulgac, Phys. Rev. B {\bf 48}, 2721 (1993);
M. Fosmire and A. Bulgac, {\em ibid.} {\bf 52}, 17509 (1995);
J. M. Thompson and A. Bulgac, {\em ibid.} {\bf 40}, 462 (1997);
L. J. Lewis, P. Jensen, and J. L. Barrat,
{\em ibid.} {\bf 56}, 2248 (1997);
S. K. Nayak, S. N. Khanna, B. K. Rao, and P. Jena,
J. Phys: Condens. Matter {\bf 10}, 10853 (1998).
F.~Calvo and P.~Labastie, 
J.~Phys.~Chem.~B {\bf 102}, 2051 (1998);
J. P. K. Doye and D. J. Wales,
Phys. Rev. B {\bf 59}, 2292 (1999);
J. Chem. Phys. {\bf 111}, 11070 (1999).
}
\bibitem{Bla97}{
P.~Blaise, S.~A.~Blundell, and C.~Guet,
Phys.~Rev.~B {\bf 55}, 15856 (1997).
}
\bibitem{Ryt98}{
A.~Rytk\"onen, H.~H\"akkinen, and M.~Manninen,
Phys.~Rev.~Lett. {\bf 80}, 3940 (1998).
}
\bibitem{Cle98}{
C. L. Cleveland, W. D. Luedtke, and U. Landman,
Phys. Rev. Lett. {\bf 81}, 2036 (1998);
Phys. Rev. B {\bf 60}, 5065 (1999). 
}
\bibitem{Cal99}{
F. Calvo and F. Spiegelmann,
Phys. Rev. Lett. {\bf 82}, 2270 (1999);
J. Chem. Phys. {\bf 112}, 2888 (2000).
}
\bibitem{Agu99a}{
A. Aguado, J. M. L\'opez, J. A. Alonso, and M. J. Stott,
J. Chem. Phys. {\bf 111}, 6026 (1999);
}
\bibitem{Agu99b}{
J. Phys. Chem. B, submitted (preprint available at http://xxx.lanl.gov/abs/physics/9911042).
}
\bibitem{Mar96}{
T.~P.~Martin,
Phys.~Rep. {\bf 273}, 199 (1996).
}
\bibitem{Sch97}{
M.~Schmidt, R.~Kusche, W.~Kronm\"uller, B.~von Issendorff, and H.~Haberland, 
Phys.~Rev.~Lett. {\bf 79}, 99 (1997);
M.~Schmidt, R.~Kusche, B.~von Issendorff, and H.~Haberland,
Nature {\bf 393}, 238 (1998);
R. Kusche, Th. Hippler, M. Schmidt, B. von Issendorff, and H. Haberland,
Eur. Phys. J. D {\bf 9}, 1 (1999).
}
\bibitem{Sch99}{
M. Schmidt, C. Ellert, W. Kronm\"uller, and H. Haberland,
Phys. Rev. B {\bf 59}, 10970 (1999).
}
\bibitem{Hab99}{
H. Haberland, in ``{\em Metal Clusters}'', ed. W. Ekardt (John Wiley $\&$ Sons,
1999), p. 181.
} 
\bibitem{Jel86}{
J.~Jellinek, T.~L.~Beck, and R.~S.~Berry,
J.~Chem.~Phys. {\bf 84}, 2783 (1986).
}
\bibitem{Guv93}{
Z.~B.~G\"uvenc and J.~Jellinek, 
Z. Phys. D {\bf 26}, 304 (1993).
}
\bibitem{Che96}{
V.~K.~W.~ Cheng, J.~P.~Rose, and R.~S.~Berry,
Surf.~Rev.~Lett. {\bf 3}, 347 (1996).
}
\bibitem{Che91}{
H. P. Cheng and R.~S.~Berry,
Phys. Rev. A {\bf 45}, 7969 (1991);
R. S. Berry, in {\em Clusters of Atoms and Molecules}, edited by
H. Haberland (Springer, Berlin, 1994), pp. 187--204.
}
\bibitem{Pea93}{
M.~Pearson, E.~Smargiassi, and P.~A.~Madden,
J.~Phys.:~Condens.~Matter {\bf 5}, 3221 (1993).
}
\bibitem{Car85}{
R.~Car and M.~Parrinello,
Phys.~Rev.~Lett. {\bf 55}, 2471 (1985);
M. C. Payne, M. P. Teter, D. C. Allan, T. A. Arias, and J. D. Joannopoulos,
Rev. Mod. Phys. {\bf 64}, 1045 (1992).
}
\bibitem{Hoh64}{
P.~Hohenberg and W.~Kohn,
Phys.~Rev. {\bf 136}, 864B (1964).
}
\bibitem{Koh65}{
W.~Kohn and L.~J.~Sham,
Phys.~Rev. {\bf 140}, 1133A (1965).
}
\bibitem{Sma94}{
E.~Smargiassi and P.~A.~Madden,
Phys.~Rev.~B {\bf 49}, 5220 (1994);
M.~Foley, E.~Smargiassi, and P.~A.~Madden,
J.~Phys.:~Condens.~Matter {\bf 6}, 5231 (1994);
E.~Smargiassi and P.~A.~Madden,
Phys.~Rev.~B {\bf 51}, 117 (1995);
{\em ibid.} {\bf 51}, 129 (1995);
M.~Foley and P.~A.~Madden,
{\em ibid.} {\bf 53}, 10589 (1996);
B.~J.~Jesson, M.~Foley, and P.~A.~Madden,
{\em ibid.} {\bf 55}, 4941 (1997);
J. A. Anta, B. J. Jesson, and P. A. Madden,
{\em ibid.} {\bf 58}, 6124 (1998).
}
\bibitem{Gov99}{
N. Govind, Y. A. Wang, and E. A. Carter,
J. Chem. Phys. {\bf 110}, 7677 (1999).
}
\bibitem{Sha94}{
V.~Shah, D.~Nehete, and D.~G.~Kanhere,
J.~Phys.:~Condens.~Matter {\bf 6}, 10773 (1994);
D.~Nehete, V.~Shah, and D.~G.~Kanhere,
Phys.~Rev.~B {\bf 53}, 2126 (1996);
V.~Shah and D.~G.~Kanhere,
J.~Phys.:~Condens.~Matter {\bf 8}, L253 (1996);
V.~Shah, D.~G.~Kanhere, C.~Majumber, and G.~P.~Das,
{\em ibid.} {\bf 9}, 2165 (1997);
A.~Vichare and D.~G.~Kanhere,
J.~Phys.:~Condens.~Matter {\bf 10}, 3309 (1998);
A.~Vichare and D.~G.~Kanhere,
Eur. Phys. J. D {\bf 4}, 89 (1998);
A. Dhavale, V. Shah, and D. G. Kanhere,
Phys. Rev. A {\bf 57}, 4522 (1998).
}
\bibitem{Gov95}{
N.~Govind, J.~L.~Mozos, and H.~Guo,
Phys.~Rev.~B {\bf 51}, 7101 (1995);
Y. A. Wang, N. Govind, and E. A. Carter,
{\em ibid.} {\bf 58}, 13465 (1998).
}
\bibitem{Bon97}{
V.~Bonaci\'c-Kouteck\'y, J.~Jellinek, M.~Wiechert, and P.~Fantucci,
J.~Chem.~Phys.~ {\bf 107}, 6321 (1997);
D. Reichardt, V.~Bonaci\'c-Kouteck\'y, P.~Fantucci, and J.~Jellinek,
Chem. Phys. Lett. {\bf 279}, 129 (1997).
}
\bibitem{Kum99}{
S. K\"ummel, P. -G. Reinhard, and M. Brack,
Eur. Phys. J. D {\bf 9}, 149 (1999);
S. K\"ummel, M. Brack, and P. -G. Reinhard,
unpublished results.
}
\bibitem{Gar98}{
I. L. Garz\'on, K. Michaelian, M. R. Beltr\'an, A. Posada-Amarillas, P. Ordej\'on,
E. Artacho, D. S\'anchez-Portal, and J. M. Soler,
Phys. Rev. Lett. {\bf 81}, 1600 (1998).
}
\bibitem{Sol20}{
J. M. Soler, M. R. Beltr\'an, K. Michaelian, I. L. Garz\'on, P. Ordej\'on,
D. S\'anchez-Portal, and E. Artacho,
Phys. Rev. B {\bf 61}, 5771 (2000).
}
\bibitem{Doy96}{
J. P. K. Doye and D. J. Wales,
J. Phys. B {\bf 29}, 4859 (1996).
}
\bibitem{Mar83}{
{\em Theory of the inhomogeneous electron gas.} Editors S. Lundqvist and N. H.
March. Plenum Press, New York (1983).
}
\bibitem{Yan86}{
W.~Yang,
Phys.~Rev.~A {\bf 34}, 4575 (1986).
}
\bibitem{Per92}{
J.~P.~Perdew,
Phys.~Lett.~A {\bf 165}, 79 (1992).
}
\bibitem{Per81}{
J.~P.~Perdew and A.~Zunger,
Phys.~Rev.~B {\bf 23}, 5048 (1981).
}
\bibitem{Cep80}{
D.~Ceperley and B.~Alder,
Phys.~Rev.~Lett. {\bf 45}, 566 (1980).
}
\bibitem{Fio95}{
C.~Fiolhais, J.~P.~Perdew, S.~Q.~Armster, J.~M.~McLaren, and H.~Brajczewska,
Phys.~Rev.~B {\bf 51}, 14001 (1995);
{\em ibid.} {\bf 53}, 13193 (1996).
}
\bibitem{Nog96}{
F.~Nogueira, C.~Fiolhais, J.~He, J.~P.~Perdew, and A.~Rubio,
J.~Phys.:~Condens.~Matter {\bf 8}, 287 (1996).
}
\bibitem{Ver65}{
L.~Verlet,
Phys.~Rev. {\bf 159}, 98 (1967);
W.~C.~Swope and H.~C.~Andersen,
J.~Chem.~Phys. {\bf 76}, 637 (1982).
}
\bibitem{Sug91}{
S.~Sugano,
{\em Microcluster Physics}, Springer-Verlag, Berlin (1991).
}
\bibitem{Erc91}{
F. Ercolessi, W. Andreoni, and E. Tosatti,
Phys. Rev. Lett. {\bf 66}, 911 (1991).
}
\end{references}
\end{document}